\documentclass[%
amsmath,amssymb,
]{revtex4-1}

\usepackage{graphicx}
\usepackage{mathtools}
\usepackage[utf8]{inputenc}
\usepackage{ulem}
\usepackage{subcaption}
\usepackage{float}
\usepackage{tabu}
\usepackage{url}
\usepackage{dcolumn}
\usepackage{bm}

\usepackage[percent]{overpic}
\usepackage{pict2e} 
\usepackage{color}

\pdfoutput=1

\begin{document}

\title{Bistability in Rayleigh-Bénard convection with a melting boundary}

\author{J. Purseed} 
\affiliation{Aix-Marseille University, CNRS, Centrale Marseille, IRPH\'E, Marseille, France}
\author{E. W. Hester}
\affiliation{University of Sydney School of Mathematics and Statistics, Sydney, NSW 2006, Australia}
\author{B. Favier} \altaffiliation[For correspondence : ]{favier@irphe.univ-mrs.fr}
\affiliation{Aix-Marseille University, CNRS, Centrale Marseille, IRPH\'E, Marseille, France}
\author{L. Duchemin}
\affiliation{Aix-Marseille University, CNRS, Centrale Marseille, IRPH\'E, Marseille, France}

\date{\today}

\begin{abstract}
A pure and incompressible material is confined between two plates such that it is heated from below and cooled from above.
When its melting temperature is comprised between these two imposed temperatures, an interface separating liquid and solid phases appears.
Depending on the initial conditions, freezing or melting occurs until the interface eventually converges towards a stationary state.
This evolution is studied numerically in a two-dimensional configuration using a phase-field method coupled with the Navier-Stokes equations.
Varying the control parameters of the model, we exhibit two types of equilibria: diffusive and convective.
In the latter case, Rayleigh-B\'enard convection in the liquid phase shapes the solid-liquid front, and a macroscopic topography is observed.
A simple way of predicting these equilibrium positions is discussed and then compared with the numerical simulations.
In some parameter regimes, we show that multiple equilibria can coexist depending on the initial conditions.
We also demonstrate that, in this bi-stable regime, transitioning from the diffusive to the convective equilibrium is inherently a nonlinear mechanism involving finite amplitude perturbations. 
\end{abstract}

\maketitle

\section{\label{sec:level1}Introduction}

Many geological patterns result from the interaction between a fluid flow and a solid front~\citep{Meakinrspa20090189}.
Erosion is one such example where the shear stress exerted by the flow can sculpt an erodible body~\citep{ristroph2012sculpting, moore2013self}.
It also plays a role in the smoothing of sharp edges~\citep{domokos2014river} and is essential in geological dating, for example, the inference of water on Mars due to erosion channels and river islands~\citep{malin2000evidence,baker2001water}.
Solid-liquid phase transition is another way of obtaining a growing interface and these transitions usually fall in the Stefan problems category with a well-defined dynamical interface separating the two phases~\citep{huppert1990fluid, worster2000solidification}. Whether it is due to dissolution or melting, the combination of a phase-change and fluid motion can lead to non-trivial topographies, for example, scalloped icebergs as a result of oceanic flow~\citep{claudin_duran_andreotti_2017, ristroph_2018} or natural shaping of dissolvable bodies or ice spheres in imposed flows~\citep{hao2001melting,hao2002heat,machicoane2013,mac2015shape}. 
The characteristics of the flow involved in such problems and the material properties of the solid phase can affect the shape of the solid/liquid front. 
For instance, an imposed flow of a binary alloy along its solid phase can suppress morphological instabilities, or trigger travelling waves\citep{delves1968theory,delves1971theory,Jiang_2013}. 
In a similar fashion, Gilpin et al. \citep{gilpin1980wave} studied  experimentally the interaction between a warm turbulent flow and an ice-water front.
If a local perturbation on the ice-water surface is added initially, an interfacial instability grows in the form of a rippled surface. 

Another interesting configuration arises when the flow is not imposed externally but is instead buoyancy-driven. 
This natural mechanism is known to generate complex topographies, as a consequence of non-uniform convective heat fluxes that cause local melting or freezing.
Applications of this convection/melting coupling are numerous.
It has significant impact on the understanding of the Earth's inner core solidification in the presence of the convecting outer liquid core~\citep{Alboussiere2010, Labrosse2017};  it affects the thermal evolution of magma oceans~\citep{Ulvrova}, provides insight on the melting of ice shelves~\citep{Martin1977,silvano2018freshening}.
This coupling also finds its application in astrophysical bodies such as Europa or Enceladus in an attempt to understand the eruption of water from the icy surface~\citep{manga2007pressurized}, trapped water bodies~\citep{walker2015ice} or the global shape of the ice-water surface and thickness of the ice crust.
In the industry, solidification of liquid metal in complex moulds~\citep{cervera1999thermo} often gives rise to natural convection; which can affect dendrites formation during crystal growth \citep{glicksman1986interaction,BECKERMANN1999468}.
In all of these examples, from large-scale geophysical applications to small-scale industrial processes, the main challenge lies in the complex dynamics of the interface between the solid and liquid phases, which depend on the imbalance between the convective and diffusive heat fluxes on both sides of the interface.

The interaction between a convective flow and a melting solid has recently received some attention~\citep{Vasil2011,esfahani2018basal,favier_purseed_duchemin_2019} where the gradual melting of a pure isothermal solid is investigated considering a standard Rayleigh-B\'enard configuration.
The melting process causes a vertical growth of the liquid layer until the critical height is reached and convective instabilities set in.
The numerical study by \cite{favier_purseed_duchemin_2019} shows that, as the convection cells are stretched, due to the vertical growth of the solid-liquid boundary, convection cells merge creating wider ones, thus respecting the aspect ratio one would observe in classical Rayleigh-B\'enard convection~\citep{Chandra1961}.
During this slow evolution, the convective heat flux has been showed to be consistent in first approximation with that of classical Rayleigh-B\'enard \cite{esfahani2018basal,favier_purseed_duchemin_2019}.
The case where a material, confined between two horizontal boundaries, is heated from below and cooled from above has also been studied experimentally by Davis {\it et. al}~\citep{davis_muller_dietsche_1984}.
They investigate the effects of the solid thickness on the upper boundary on the onset of convection and showed that the critical Rayleigh number is significantly reduced.
A weakly non-linear analysis was also performed and they found that convection was still possible below the convective instability threshold and that the bifurcation becomes trans-critical. Their findings were then verified experimentally and bistable states were reported close to the instability threshold \citep{dietsche1985influence}.
A detailed description of the equilibrium states in such a system, close and far for the convective instability threshold, remains however to be studied, which is the main motivation of this paper.

In the present paper, a configuration similar to \cite{davis_muller_dietsche_1984} is numerically studied where the melting temperature and the temperature difference between the two plates are free parameters and are varied in an attempt to find an equilibrium.
We are also interested in the dynamics and the stability of these equilibria.
The paper is structured as follows~: we give a general formulation of the physical setup in section \ref{sec:level2}, followed by a description of the equilibrium states theoretically and their comparison to the numerical simulations in section \ref{sec:level3}.
The existence of a bi-stability regime is discussed in section \ref{sec:level4}.
We finally conclude in section \ref{sec:level10}. 
A brief description of the numerical method is given in Appendix~\ref{sec:level11} which is identical to the one proposed by \cite{favier_purseed_duchemin_2019} and thus for a more detailed description, interested readers are referred to that particular paper. 
\section{\label{sec:level2}Mathematical model}

\begin{figure*}[t!]
\centering
    \begin{subfigure}[b]{0.48\textwidth}
         \centering
         \includegraphics[width=1.0\linewidth]{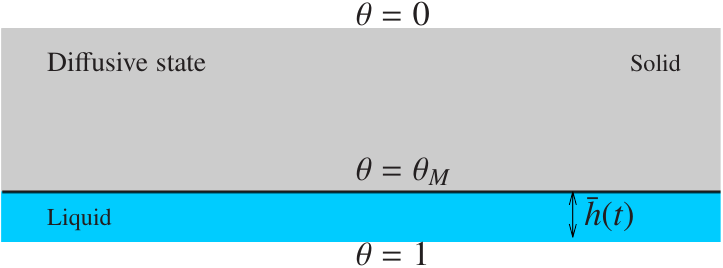}
         \caption{}
    \end{subfigure}
     \hfill
     \begin{subfigure}[b]{0.48\textwidth}
         \centering
         \includegraphics[width=1.0\linewidth]{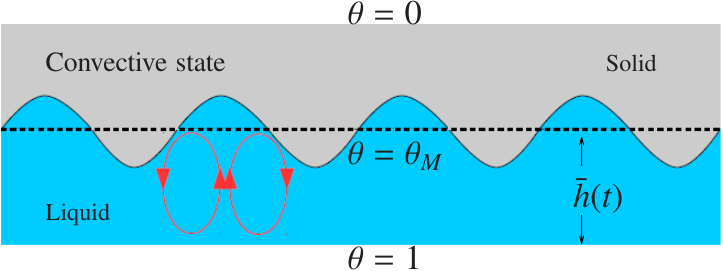}\\
         \caption{}
     \end{subfigure}
\caption{Schematic representation of the problem. The solid phase is shown in grey while the liquid is in blue.
On the left, heat transfer is entirely due to conduction through the whole layer.
On the right, convection is the predominant source of heat transfer in the liquid.
The thick black line represents the solid-liquid interface which corresponds to a dimensionless temperature of $\theta_{M}$.}
\label{fig:schema}
\end{figure*}

Our idealised problem is represented in Figure~\ref{fig:schema}, where we bound a solid and its corresponding liquid phase by two horizontal walls while the system is two-dimensional (2D) and periodic in the horizontal direction.
The two rigid horizontal plates are separated by a distance $H$ while the horizontal extent of the periodic domain is $\lambda H$ with $\lambda$ the aspect ratio.
The imposed temperature of the bottom plate is $T_1$, the temperature of the top plate is $T_0$, and the melting temperature $T_M$ is such that $T_0<T_M<T_1$.
Both plates are assumed to be impenetrable and no-slip.
The physical properties of both solid and liquid phases are assumed to be constant and equal.
The thermal diffusive time $H^2/\kappa$ is used as a reference for the time scale, $\kappa$ being the constant thermal diffusivity.
$H$ is used as the reference length and $\Delta T = T_{1}-T_{0}$ is the temperature scale.
The governing dimensionless equations in the Boussinesq approximation for the fluid phase are given by
\begin{align}
\label{eq:momentum}
\frac{1}{Pr}\left(\frac{\partial\bm{u}}{\partial t}+\bm{u}\cdot\nabla\bm{u}\right) & =-\nabla P+ Ra \; \theta \; \bm{e}_z+\nabla^2\bm{u} \\
\frac{\partial \theta}{\partial t}+\bm{u}\cdot\nabla \theta & =\nabla^2\theta \\
\label{eq:div}
\nabla\cdot\bm{u} & =0
\end{align}
where $\theta=(T-T_{0})/(T_{1}-T_{0})$ is the dimensionless temperature, $\bm{u}=(u,w)$ is the two-dimensional velocity field, $Ra$ is the Rayleigh number based on the total height $H$ and $Pr$ is the usual Prandtl number 
\begin{equation}\label{equ:Rayleigh_Thermal_Expansion}
Ra=\frac{\alpha g \Delta TH^3}{\nu\kappa} \quad \textrm{and} \quad Pr=\frac{\nu}{\kappa},
\end{equation}
where $\alpha$ is the thermal expansion coefficient, $g$ is the gravitational acceleration and $\nu$ the kinematic viscosity.
Note that by analogy with standard Rayleigh-B\'enard configurations, we choose the global temperature difference as a reference.
For simplicity, in the whole study, the Prandtl number is taken to be one and only the Rayleigh number is varied.
The solid phase is considered to be non-deformable and stationary ($\bm{u}=\bm{0}$) and accordingly, we need only to solve the dimensionless heat equation
\begin{equation}\label{equ:heat_equation}
    \frac{\partial \theta}{\partial t} = \nabla^{2} \theta \ .
\end{equation}
In comparison to the classical Rayleigh-B\'enard convection problem, for which there is only a liquid phase, a dynamical phase-change boundary separates the liquid and the solid.
These internal boundary conditions are given by the Stefan conditions \citep{huppert1990fluid} 
\begin{align}
\label{eq:st1}
\theta & = \theta_M\\
\label{eq:st2}
St \ \bm{v}\cdot\bm{n} & =\left(\nabla \theta^{(S)}-\nabla \theta^{(L)}\right) \cdot \bm{n} \ ,
\end{align}
where $\theta_M=\left(T_{M}-T_{0})/(T_{1}-T_{0}\right) \in ]0,1[$ is the dimensionless melting temperature, $\bm{n}$ is the normal to the interface pointing towards the liquid phase, $\bm{v}$ is the velocity of the interface and superscripts $(S)$ and $(L)$ denote solid and liquid phases respectively.
The Stefan number $St$ corresponds to a dimensionless ratio between the latent heat associated with the solid-liquid transition $\mathcal{L}$ and the characteristic specific heat of the system
\begin{equation}
\label{eq:stefan}
St = \frac{\mathcal{L}}{c_p\Delta T} \ ,
\end{equation}
where $c_p$ is the heat capacity at constant pressure.
Equation~\eqref{eq:st2} expresses the fact that the interface moves with a normal velocity proportional to the heat-flux jump across the interface.
A steady interface therefore corresponds to a balance between the heat fluxes across it.
We assume the same density for the two phases so that the interface is considered to be impenetrable and no-slip boundary conditions are applied to it \citep{davis_muller_dietsche_1984}.
The Gibbs-Thomson effect due to the surface energy of the solid-liquid interface is neglected \citep{worster2000solidification}.
This thermodynamical effect is nevertheless the starting point when deriving a diffuse-interface method called the phase-field method \citep{BECKERMANN1999468}.
The problem described above is solved numerically by using a mixed pseudo-spectral fourth-order finite-difference method \citep{favier2014,favier_guervilly_knobloch_2019} and the particular phase-field model which has been discussed and validated in \citep{favier_purseed_duchemin_2019}.
For several cases, we also checked our results by using the open-source pseudo-spectral solver Dedalus \citep{DEDALUS,burns2019} (more information at \url{http://dedalus-project.org}). 
More details about the model equations and numerical parameters are given in Appendix~\ref{sec:level11}.

\section{\label{sec:level3}Equilibrium states}

The case of a nearly isothermal solid, discussed in \cite{esfahani2018basal} and \cite{favier_purseed_duchemin_2019}, leads to a complete melting of the solid phase until the upper boundary is reached.
Following these studies, we turn our interest to the case for which the temperature of the upper plate is fixed and lower than the melting temperature.
In this configuration, we expect equilibrium states for which the heat flux in the solid is statistically balanced by the heat flux in the liquid, consistently with equation~\eqref{eq:st2}.
Hence, this section is dedicated to predicting the average fluid depth at equilibrium by balancing the average heat flux in both phase and comparing this prediction to numerical simulations.

The following configuration is chosen for all the simulations~: the initial position of the interface is set to $z=h_{0}=0.1$ (where $z=0$ corresponds to the bottom plate), and the horizontal length of the numerical domain is set to $\lambda=6$, in order to avoid any confinement phenomenon.
For simplicity, both the Prandtl number and the Stefan number are fixed to unity.
The simulations are initialised with a fluid at rest and a piece-wise linear temperature profile given by
\begin{align}
\label{eq:tinit}
    \theta(t=0) & =   \left\{
      \begin{aligned}
        1+\left(\theta_M-1\right)z/h_0 & \quad \textrm{if} \quad z\le h_0\\
        \theta_M\left(z-1\right)/\left(h_0-1\right) & \quad \textrm{if} \quad z>h_0 \ .
      \end{aligned}
    \right.
\end{align}
This initial condition is not generally at equilibrium since there is a heat flux discontinuity at $z=h_0$.
We then add small amplitude temperature perturbations in the liquid phase in order to potentially trigger the Rayleigh-B\'enard instability.
Starting from this initial condition, the computations always reach a stationary state, which can be described according to the asymptotic value $h_\infty$ of the mean height of the fluid-solid interface
\begin{equation}
\label{eq:meanh}
    \overline{h} = \frac{1}{\lambda} \int^{\lambda}_{0} h\left(x,t\right) \text{d}x,
\end{equation}
where $\lambda$ is the dimensionless length of our domain and $h\left(x,t\right)$ is the local vertical position of the interface (found by computing the contour $\phi=1/2$ or equivalently $\theta=\theta_M$). 
This equilibrium state is assumed to be reached when the average kinetic energy in the liquid and the averaged height $\overline{h}$ are constant over time, which is typically the case after several thermal diffusion times.
This protocol is repeated for multiple melting temperatures ranging from $\theta_M=0.1$ to $0.9$ and for multiple Rayleigh numbers from $Ra=10^{4}$ to $10^{7}$.
The asymptotic value $h_\infty$ is represented in Figure~\ref{fig:Main_figure}(a) for all the computations, as a function of $1-\theta_M$ and for different input Rayleigh numbers $Ra$.
Two types of equilibria are observed and discussed in the following~: diffusive and convective equilibria. 

\subsection{\label{sec:no_convection}Diffusive equilibria}

\begin{figure}
     \centering
         \begin{subfigure}[c]{0.48\textwidth}
         \centering
         \includegraphics[]{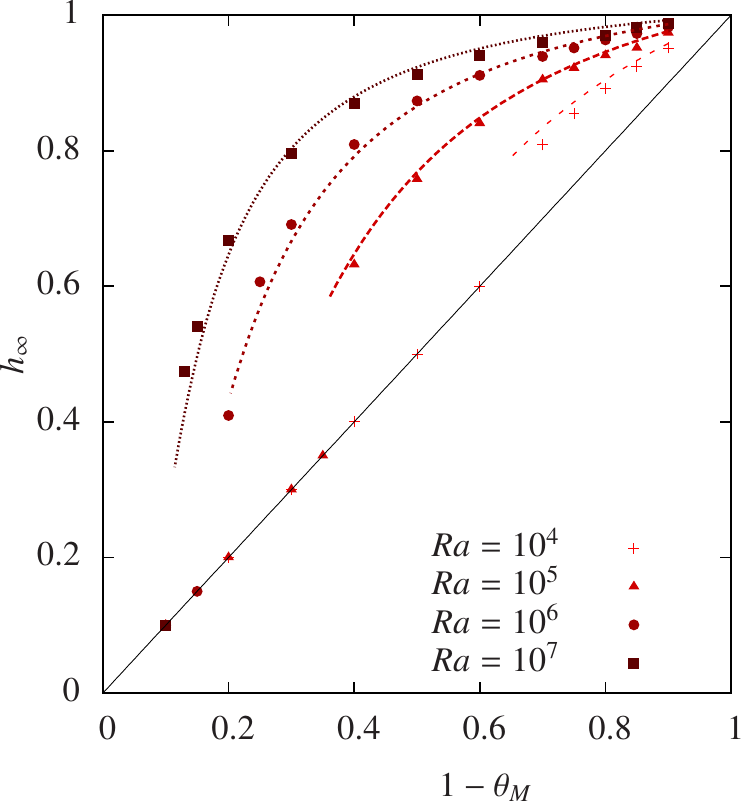}
                  \caption{}
                  \label{fig:fig2a}
         \end{subfigure}
        \begin{subfigure}[rt]{0.48\textwidth}
         \centering
         \includegraphics[]{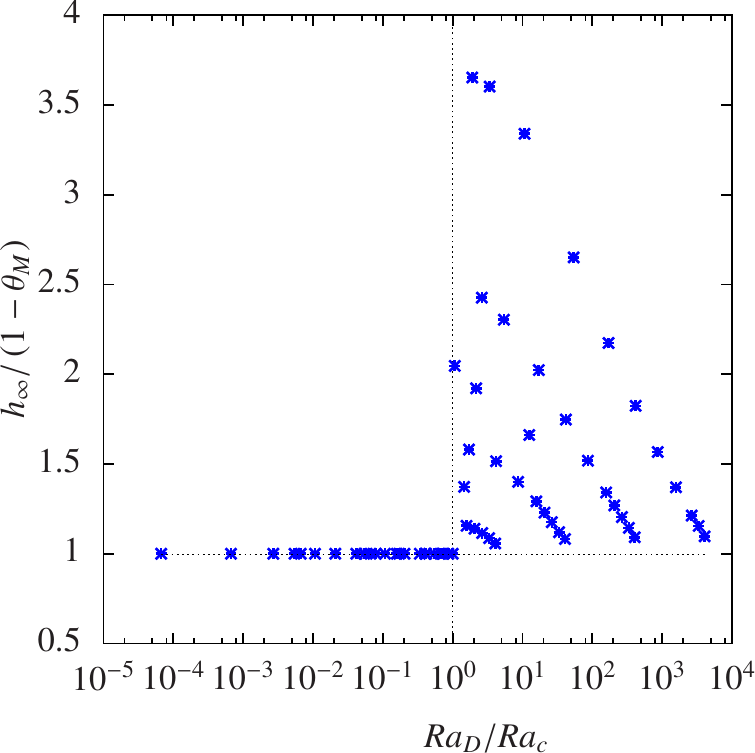}
                  \caption{}
                  \label{fig:fig2b}
     \end{subfigure}
        \caption{(a) Equilibrium height of the interface as a function of the melting temperature for different Rayleigh numbers. Dashed curves are obtained by equating the fluxes across the solid and liquid layers and estimating the Nusselt number following equation~\eqref{eq:Nusselt_scaling}. (b) Equilibrium height normalised by the diffusive equilibrium height \eqref{eq_diff} as a function of a normalised Rayleigh effective number.}
       \label{fig:Main_figure}
\end{figure}

In some of our computations, an equilibrium is reached without observing any motion inside the liquid phase~: this equilibrium is purely diffusive. 
In this case, the stationary state is fully described by the steady solution of the heat equation~\eqref{equ:heat_equation} in both phases leading to
\begin{equation}
    \theta = 1 - z \qquad \mathrm{and} \qquad h_\infty = 1 - \theta_M \ . 
    \label{eq_diff}
\end{equation}
The (diffusive) heat-fluxes in the solid and liquid phases are then equal and given respectively by
\begin{align}
\label{eq:heatflux_solid_diff}
Q_{D}^{(S)}  & = \frac{\theta_{M}}{1-h_\infty} \ ,\\
\label{eq:heatflux_liquid_diff}
Q_{D}^{(L)} & = \frac{1-\theta_{M}}{h_\infty} \ .
\end{align}
When fluid motion is absent, the melting temperature $\theta_{M}$ is the only parameter which dictates the equilibrium height and the latter increases with decreasing $\theta_{M}$.
The points along the oblique straight line whose equation is \eqref{eq_diff} in figure~\ref{fig:Main_figure}(a) represent computations showing this kind of equilibrium states.

\subsection{Convection onset}

As we vary the Rayleigh number and the melting temperature, some simulations depart from the diffusive base state described by equation~\eqref{eq_diff}.
These cases are all characterised by fluid motions in the form of convective rolls and non-planar phase-change interface (see Figure~\ref{fig:figure4}(b) below for example).
A simple way of knowing beforehand whether the diffusive base state discussed earlier is stable or not is to define the effective Rayleigh number of the fluid layer as
\begin{equation}
Ra_e=Ra \left( 1- \theta_{M} \right) \overline h^3 \ ,
    \label{eq:Reff_definition}
\end{equation}
where $1-\theta_M$ is the effective temperature difference across the fluid layer and $\overline{h}$ is the averaged fluid depth as defined in equation~\eqref{eq:meanh}.
Note that this definition of an effective Rayleigh number is analogous to the one described by \cite{couston2017dynamics} for the case of thermal convection interacting with a stably-stratified fluid layer above.
For the diffusive state defined by equation~\eqref{eq_diff}, the fluid depth at equilibrium is simply $h_{\infty}=1-\theta_M$ and the effective Rayleigh number is then
\begin{equation}
	Ra_D=Ra \left(1-\theta_M\right)^4 \ . 
	\label{eq:effdiff}
\end{equation}

Note that the critical Rayleigh number is not the standard value of $Ra_c\approx1707$ \citep{Chandra1961} valid for fixed temperature and no-slip boundaries.
Due to the effect of heat diffusion in the adjacent solid layer, it has been showed \citep{davis_muller_dietsche_1984,toppaladoddi_wettlaufer_2019} that the critical Rayleigh number $Ra_c$ varies from 1707 for very thin solid layers (\textit{i.e.} $\overline{h}\rightarrow1$) down to approximately 1493 for thick solid layers (\textit{i.e.} $\overline{h}\rightarrow0$).
This dependence of the critical Rayleigh number on the solid layer thickness is taken into account in the following results.

For each values of $Ra$ and $\theta_M$, the effective Rayleigh number of the diffusive equilibrium $Ra_D$ can be compared with the critical Rayleigh number $Ra_c$.
Hence, for all the values greater than the critical value, the equilibrium state will be a convective one and, for all values that are smaller, one can expect a diffusive equilibrium.
This is further confirmed by figure \ref{fig:Main_figure}(b), where we show the ratio $h_\infty/(1-\theta_M)$ as a function of $Ra_D/Ra_{c}$.
We can see clearly the threshold between the diffusive and convective regimes~: below one, the equilibrium is diffusive and $h_\infty=1-\theta_M$, whereas above one, $h_\infty > 1-\theta_M$ due to the increased convective heat flux.
Note that some convective equilibria are very close to the marginal line.
We discuss into more details the behaviour close to the threshold later in section~\ref{sec:level4}.

\subsection{\label{sec:with_convection}Convective equilibria}

The challenge in describing the convective equilibrium states is to model the heat flux in the liquid, which is somehow analogous to the classical Rayleigh-Bénard convection, as shown in \cite{esfahani2018basal,favier_purseed_duchemin_2019}.
The Nusselt number is defined as the ratio between the total and the diffusive heat fluxes
\begin{equation}
\label{eq:Nusselt_def_flux}
    Nu = Q_T^{(L)} / Q_{D}^{(L)} = Q_T^{(L)} \bar h / \left( 1- \theta_{M} \right) \ .
\end{equation}
Looking for an equilibrium state, we equate the diffusive heat flux in the solid (\ref{eq:heatflux_solid_diff}) and the total heat flux in the liquid (\ref{eq:Nusselt_def_flux}) which leads to the following equation 
\begin{equation} \label{eq:fluxbalance}
    \frac{\theta_{M}}{1-h_{\infty}} = Nu\frac{1-\theta_{M}}{h_{\infty}} \ .
\end{equation}

Note that the Nusselt number is generally a function of the effective Rayleigh number, which itself is a function of the average fluid depth given by equation (\ref{eq:Reff_definition}).
Solving for $h_{\infty}$ in equation~\eqref{eq:fluxbalance} can therefore be non-trivial.
In the purely diffusive regime, we have $Nu=1$ by definition and we recover the solution given previously by equation~\eqref{eq_diff}.
In this section, we focus solely on solutions that are convective and far from the instability threshold, i.e, $Ra_{e} \gg Ra_{c}$.
The solutions of equation~\eqref{eq:fluxbalance} close to the threshold will be further discussed in section \ref{sec:discuss} where a more refined model for the Nusselt number will be given.
For now, in the supercritical limit far from the instability threshold, the relation between the Nusselt number and the effective Rayleigh number is considered to be of the classical form
\begin{equation}
\label{eq:Nusselt_scaling}
    Nu \sim \gamma Ra_{e}^{\beta} \ ,
\end{equation}
where $\gamma$ and $\beta$ are constants, extensively studied in the literature.
We recall that $Ra_e$ is the effective Rayleigh number based on the fluid depth as defined by equation~\eqref{eq:Reff_definition}.
If one considers a turbulent convection and high Rayleigh numbers, $\beta$ is approximately $1/3$\cite{Malkus1954}, whereas for more intermediate Rayleigh numbers, the exponent is around $1/4$\citep{grossmann_lohse_2000}.
In the following, we have chosen $\beta = 1/4$ and $\gamma = 0.27$ (Regime $I_l$ of \cite{grossmann_lohse_2000}), which is in good agreement with the Nusselt numbers measured from our simulations (see Figure~\ref{fig:Nusselt_Rayleigh}).
The Nusselt number is measured at the bottom boundary following
\begin{equation}
Nu=\left(-\frac{1}{\lambda}\int_{0}^{\lambda}\frac{\partial\theta}{\partial z} \Big \vert_{z=0} \text{d}x\right)/Q_D^{(L)} \ ,
\end{equation}
where $Q_D^{(L)}=(1-\theta)/\overline{h}$ is approximately the diffusive heat flux across the fluid layer (neglecting the fact that the interface is not planar, see \cite{favier_purseed_duchemin_2019} for more details).

By substituting equations \eqref{eq:Reff_definition} and \eqref{eq:Nusselt_scaling} into equation~\eqref{eq:fluxbalance}, we obtain an equation for the average fluid depth as a function of $\theta_{M}$, $Ra$, $\gamma$ and $\beta$.
This nonlinear equation can be solved for $h_\infty$ by using a bisection method, and there is a unique solution in the range $h_{\infty}\in[0,1]$.
The results are shown in figure \ref{fig:Main_figure}(a) by the four dashed curves.
Our computations are in good agreement with this theoretical prediction of the convective equilibrium height, which further confirms that convection below the phase change interface is equivalent to standard Rayleigh-B\'enard convection, at least in terms of averaged heat flux.

\section{\label{sec:level4}Bi-stability}

\subsection{Dependence on initial conditions}\label{sec:initial_cond}

In this section, we ask whether the long-time equilibria shown in figure \ref{fig:Main_figure} depend on the initial conditions, {\it i.e.} the value of $h_0$ in equation~\eqref{eq:tinit}.
We recall that the previous results were obtained using an arbitrary value of $h_0=0.1$.
We now systematically vary $h_0$ from $0.1$ to $0.9$.

\begin{figure}[!t]
\centering
        \begin{subfigure}[l]{0.48\textwidth}
         \centering
         \includegraphics[]{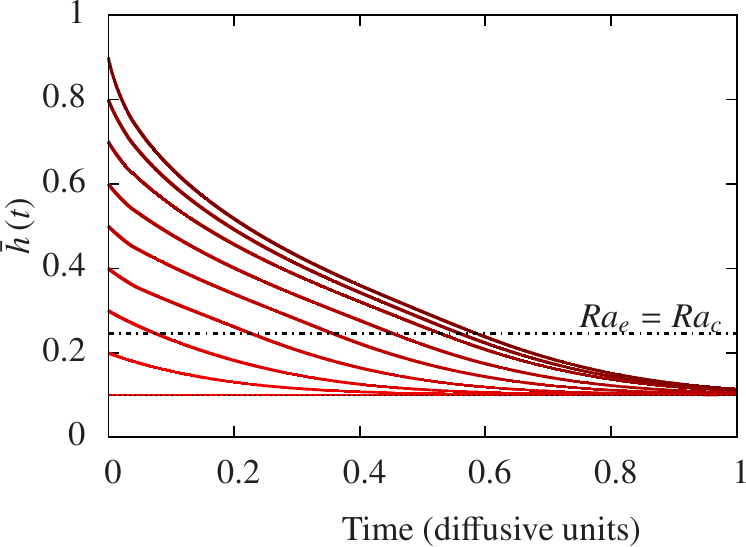}
         \caption{}
        \label{fig:RA106T09}
     \end{subfigure}
     \hfill
     \begin{subfigure}[r]{0.48\textwidth}
         \centering
         \begin{overpic}[width=8.6cm,height=3.5mm,tics=10]{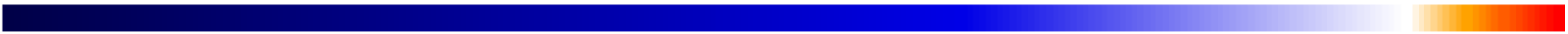}
         \put(0,4){$0$}
         \put(98,4){$1$}
         \put(87,4){$0.9$}
         \end{overpic}\\
        \vspace{2.5mm}
         \includegraphics[width=1.0\linewidth]{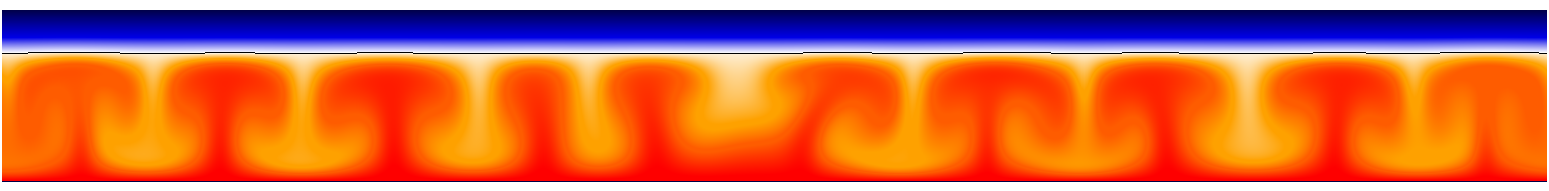}\\
         \vspace{1.5mm}
         \includegraphics[width=1.0\linewidth]{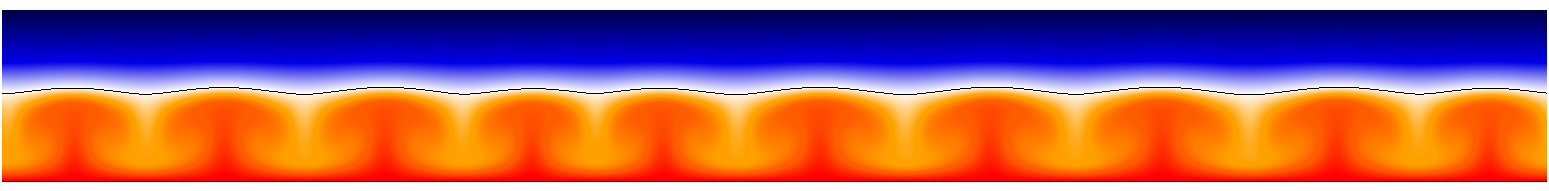}\\
         \vspace{1.5mm}
         \includegraphics[width=1.0\linewidth]{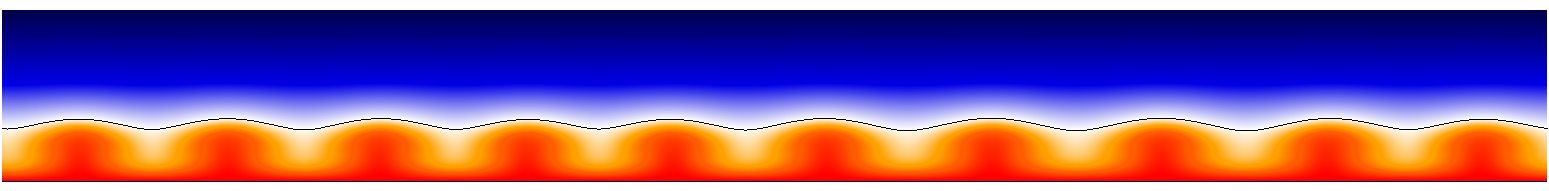}\\
         \vspace{1.5mm}
         \includegraphics[width=1.0\linewidth]{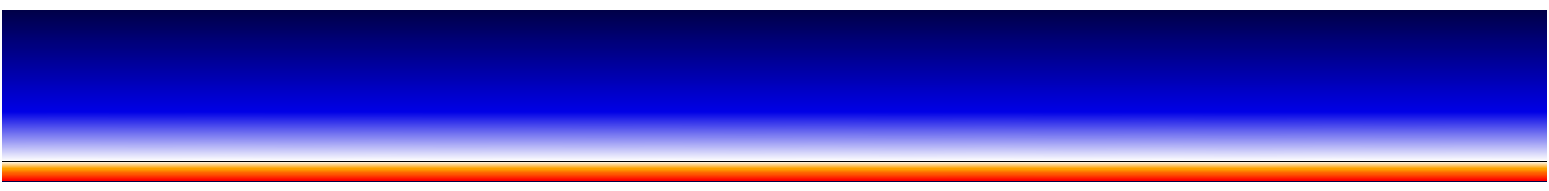}   
         \caption{}
         \label{fig:visus_fig3}
     \end{subfigure}
\caption{(a) Time evolution of the average fluid depth for multiple initial height of the interface. The parameters are $Ra=10^6$ and $\theta_M=0.9$. The horizontal dash-dott line corresponds to the critical height above which convection is sustained (\textit{i.e.} $Ra_e=Ra_c$). (b) Visualisations of the temperature field across the numerical domain are shown for the case $h_{0}=0.9$. The dark blue colour represents $\theta=0$ while the bright red colour represents $\theta=1$. The black line in the visualisation represents the solid-liquid boundary at $\theta_{M}$. Time increases from top to bottom ($t=0.04$, $0.19$, $0.44$ and $0.99$ in diffusive units).}
\label{fig:figure3}
\end{figure}

We first choose an equilibrium expected to be diffusive: $Ra=10^6$ and $\theta_M=0.9$.
For this set of parameters, the effective Rayleigh number of the diffusive equilibrium \eqref{eq:effdiff} is $Ra_D=100$ well below the critical value.
Figure \ref{fig:figure3}(a) represents the averaged fluid depth $\bar{h}(t)$ as a function of time for different initial interface positions.
All these computations converge towards $h_\infty=1-\theta_M=0.1$, which is the theoretical diffusive base state.
It is interesting to note that some of these computations present an early convection phase, which eventually disappears, eventually leading to the final diffusive equilibrium.
This is for example the case of the simulation with $h_0=0.9$, for which the initial value of the effective Rayleigh number is $Ra_e(t=0)=7.29\times10^4$, well above the critical value for the onset of convection.
The critical height (such that $Ra_{e}=Ra_{c}$) above which convection appear is represented in figure \ref{fig:figure3}(a) by the horizontal dotted line and is approximately equal to $0.2464$.
Hence, for all values of $\overline{h}$ that are greater than this critical height, convection rolls are potentially present in the liquid phase.
While this early convection slows down the solidification of the fluid layer, it is eventually overwhelmed by the dominant diffusive heat flux from the solid layer leading asymptotically to the expected diffusive equilibrium.
Such an evolution is shown in figure~\ref{fig:figure3}(b) for the case with $h_0=0.9$.

\begin{figure}[!t]
\centering
        \begin{subfigure}[l]{0.48\textwidth}
         \centering
         \includegraphics[]{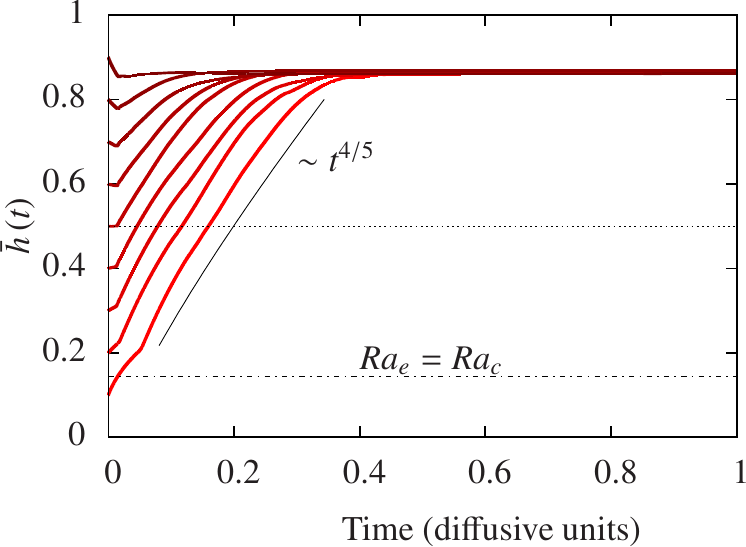}
         \caption{}
        \label{fig:RA106T05}
     \end{subfigure}
     \hfill
     \begin{subfigure}[r]{0.48\textwidth}
         \centering
         \begin{overpic}[width=8.6cm,tics=10]{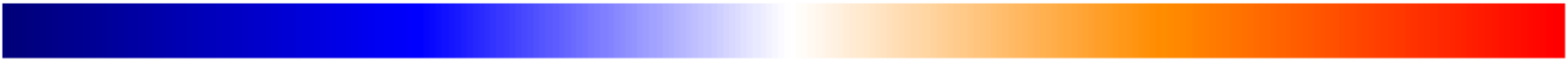}
         \put(0,5){$0$}
         \put(98,5){$1$}
         \put(48,5){$0.5$}
         \end{overpic}\\
        \vspace{2.5mm}
         \includegraphics[width=1.0\linewidth]{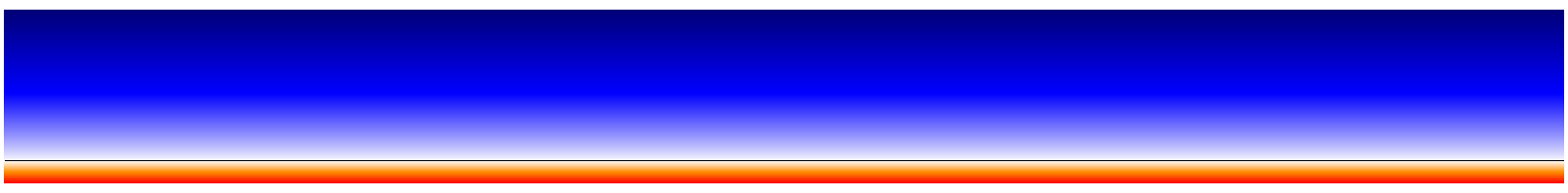}\\
         \vspace{1.5mm}
         \includegraphics[width=1.0\linewidth]{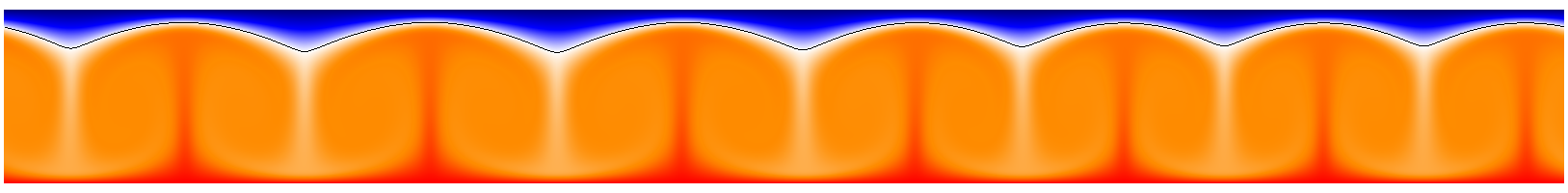}\\
         \vspace{1.5mm}
         \includegraphics[width=1.0\linewidth]{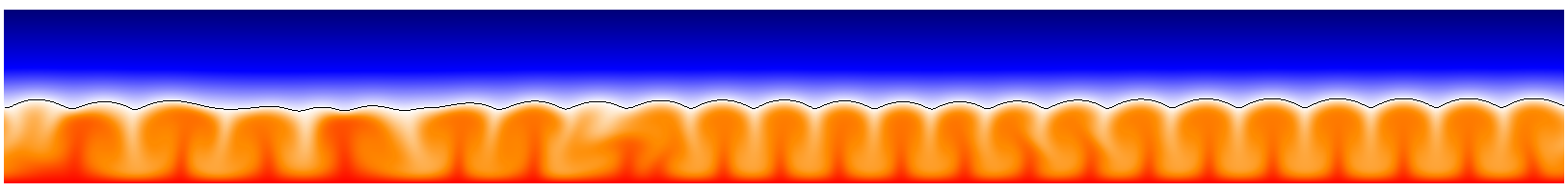}\\
         \vspace{1.5mm}
         \includegraphics[width=1.0\linewidth]{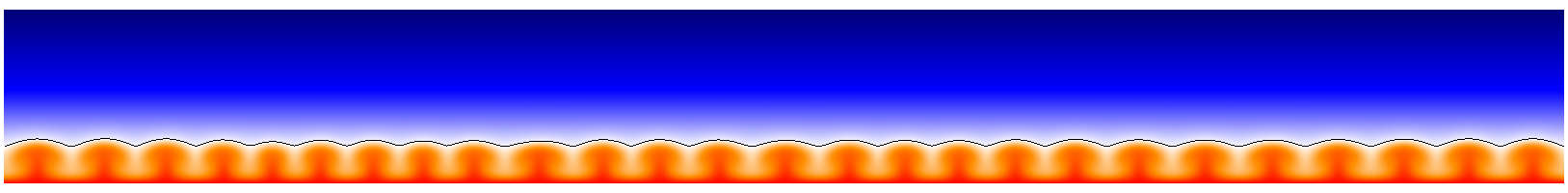}   
         \caption{}
         \label{fig:visus_fig4}
     \end{subfigure}
\caption{Same as Figure~\ref{fig:figure3} but with $Ra=10^6$ and $\theta_M=0.5$. (a) The scaling $\overline{h}\sim t^{4/5}$ predicted by \cite{favier_purseed_duchemin_2019} is shown for reference. The horizontal dotted line correspond to the diffusive equilibrium $h_{\infty}=1-\theta_M$. (b) Visualisations correspond to $t=0.01$, $0.06$, $0.14$ and $0.99$ in diffusive units.}
\label{fig:figure4}
\end{figure}

We now choose Ra = $10^6$ and $\theta_M=0.5$ which corresponds to a convective equilibrium since $Ra_D=6.25\times10^4$ is well above the onset for convection.
The temporal evolution of the interface position is represented in figure~\ref{fig:figure4}(a) for $h_0$ varying from $0.1$ to $0.9$.
As in the previous case, all simulations converge towards the same equilibrium, which is now convective as expected from the chosen parameters, and correctly predicted by equation~\eqref{eq:fluxbalance}.
Note that during the early stage of the melting, when the diffusive heat flux in the solid is negligible compared to the convective heat flux, the results of \cite{favier_purseed_duchemin_2019} are applicable.
In particular, they predicted that the averaged fluid depth should grow as $\overline{h}(t)\sim t^{1/(2-3\beta)}$ where $\beta$ is the exponent in the Nusselt scaling \eqref{eq:Nusselt_scaling}.
For our moderate Rayleigh number simulations, $\beta=1/4$ leads to $\overline{h}\sim t^{4/5}$ as observed in figure~\ref{fig:figure4}(a) at early times before the heat flux in the solid phase balances the convective heat flux.
Note that there is a slight variability in the average fluid depth at equilibrium.
This spread is due to the fact that we do not have the exact same number of convection rolls in all cases, leading to small variations in the Nusselt number.
This is a first indication that the final equilibrium state of this system depends on the initial conditions and more generally on the history of the interface.

\begin{figure}[!t]
\centering
        \begin{subfigure}[l]{0.48\textwidth}
         \centering
         \includegraphics[]{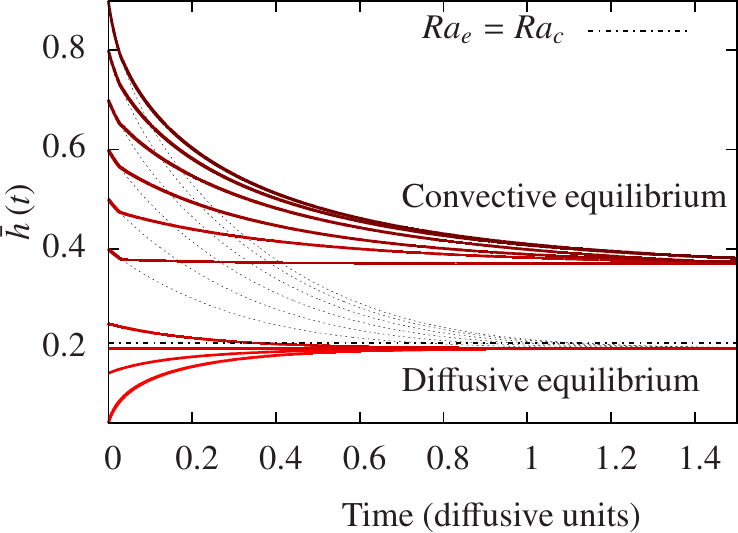}
         \caption{}
        \label{fig:bistability}
     \end{subfigure}
     \hfill
     \begin{subfigure}[r]{0.48\textwidth}
         \centering
         \begin{overpic}[width=8.6cm,tics=10]{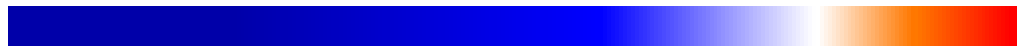}
         \put(0,6){$0$}
         \put(98,6){$1$}
         \put(77,6){$0.8$}
         \end{overpic}\\
        \vspace{4mm}
         \includegraphics[height=2cm, width=1.0\linewidth]{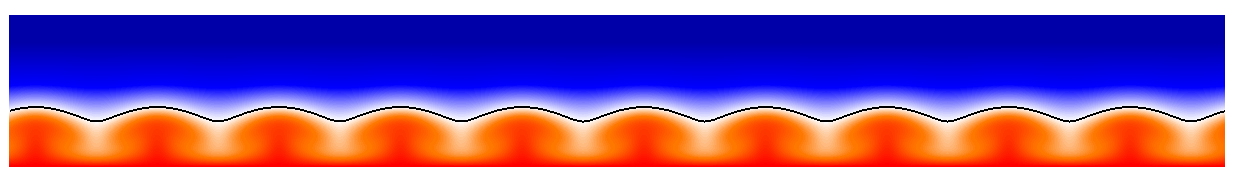}\\
         \vspace{5.5mm}
         \includegraphics[height=2cm, width=1.0\linewidth]{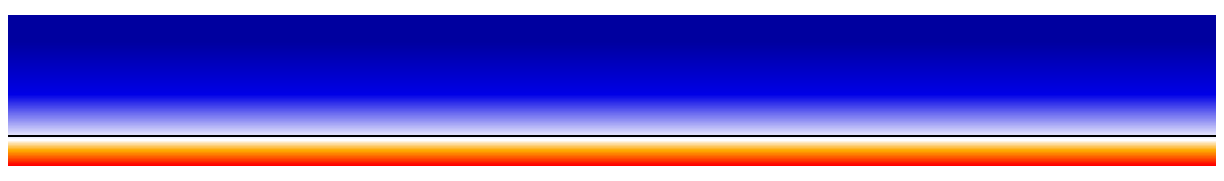}   
         \caption{}
         \label{fig:visus_fig5}
     \end{subfigure}
\caption{(a) Time evolution of the average fluid depth for multiple initial fluid depths $h_0$, $Ra=8\times10^5$, $\theta_M=0.8$ and $\lambda=8$. The dotted grey lines correspond to the purely diffusive evolution of the interface for each case (i.e. with $Ra=0$). The horizontal dash-dot black line corresponds to the critical height at which $Ra_e=Ra_c$. (b) Visualisations of the temperature field at equilibrium. The black line represents the solid-liquid boundary at $\theta=\theta_{M}=0.8$. We show both convective (top) and diffusive (bottom) solutions.}
\label{fig:figure5}
\end{figure}

At this stage, it is legitimate to wonder whether the equilibrium states are unique for a given set of parameters $\theta_M$ and $Ra$.
The two previous examples were either very stable ($Ra_D\ll Ra_c$) or very unstable ($Ra_D\gg Ra_c$) with respect to convection.
We now consider the case defined by $Ra=8\times10^5$ and $\theta_M=0.8$ for which the diffusive base state is only marginally stable with respect to convection ($Ra_D=1280$, just below the critical value which is here equal to $1493$ \citep{davis_muller_dietsche_1984}).
Figure \ref{fig:bistability} shows the evolution of $\bar{h}(t)$ as a function of time, for different values of $h_0$ ranging from $0.1$ to $0.9$, as before.
This time however, the final equilibrium is not unique and clearly depends on $h_0$.
When $h_0<0.4$, the system converges towards the expected diffusive state (since $Ra_D<Ra_c$).
More surprisingly, when $h_0>0.4$, we observe a stable convective solution even though the diffusive base state is stable for this choice of parameters.
Note that the stability of the convective solutions has been confirmed by running the simulations for at least five diffusive times.
This clearly shows that, close to the onset of convection, this system exhibits bi-stability and dependence on initial conditions. This is in agreement with the theoretical prediction of \cite{davis_muller_dietsche_1984} and the experimental observation of bi-stable states by \cite{dietsche1985influence}. We also recall that in section \ref{sec:with_convection}, we assumed that convection occurs only far from the threshold and in that limit equation (\ref{eq:Nusselt_scaling}) was used. However, for this particular case where we observe bi-stability, the convection is close to the threshold ($Ra_e\approx 5.5 \times Ra_c$).
Hence, a more refined $Nu-Ra_{e}$ scaling is required to better understand the origin of this regime, which is given later in section~\ref{sec:discuss}.

\subsection{Finite amplitude perturbations}

\begin{figure}[t!]
    \centering
    \includegraphics[]{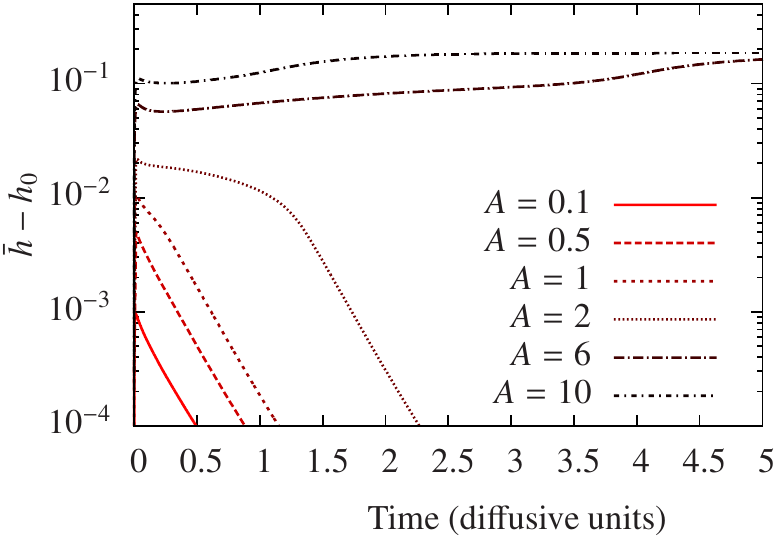}
    \caption{Time evolution of the average distance of the interface from its initial position for $\theta_{M} = 0.8$ and for multiple perturbation amplitudes. The Rayleigh number is fixed at $8\times10^{5}$.}
    \label{fig:Figure_subcritical_bifurcation}
\end{figure}

\begin{figure}[!t]
\centering
        \begin{subfigure}[l]{0.48\textwidth}
         \centering
         \includegraphics[width=0.99\textwidth]{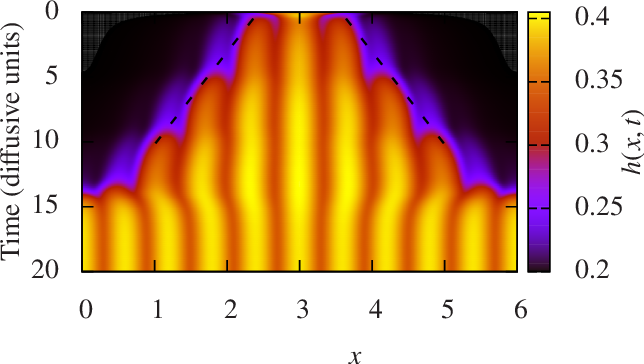}
         \caption{}
        \label{fig:ht_plot_subcritical}
     \end{subfigure}
     \hfill
     \begin{subfigure}[r]{0.48\textwidth}
         \centering
         \includegraphics[width=1.0\linewidth]{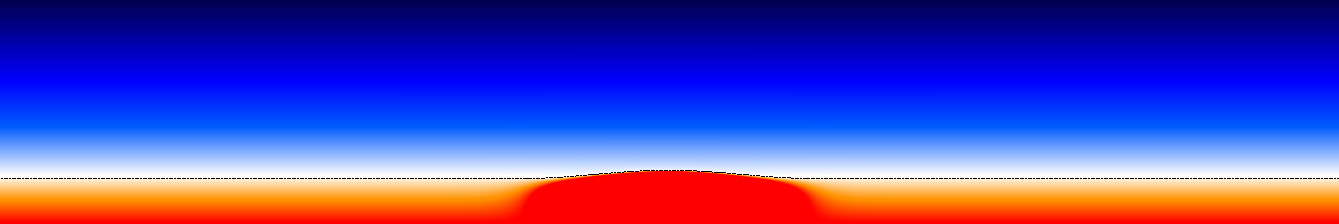}\\
         \vspace{3mm}
         \includegraphics[width=1.0\linewidth]{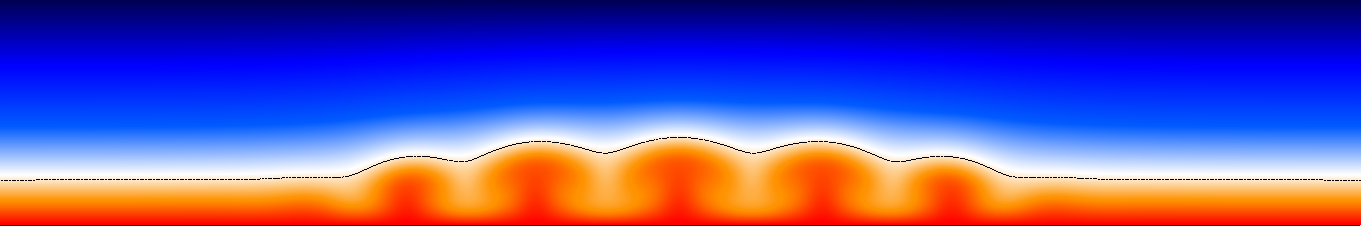}\\
         \vspace{3mm}
         \includegraphics[width=1.0\linewidth]{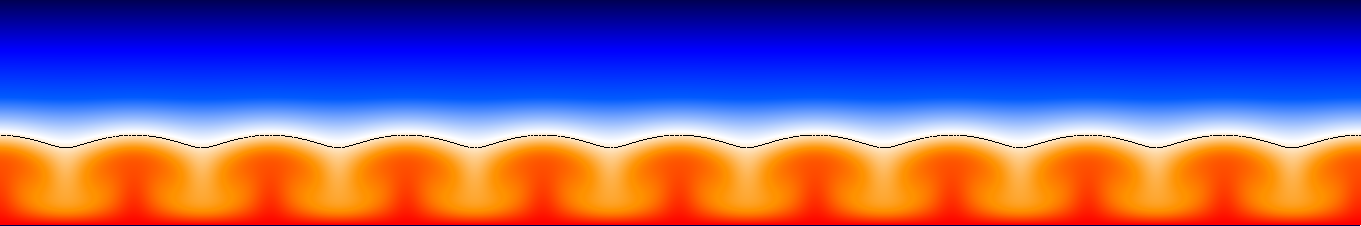}
         \caption{}
         \label{fig:Visualisation}
     \end{subfigure}
\caption{(a) Spatio-temporal evolution of the interface position for $A=6$ and $h_0=1-\theta_M=0.2$. (b) Visualisation of the temperature field showed as time is increasing from top to bottom. The dark blue colour represents $\theta=0$ while the bright red colour represents $\theta=1$. The interface, $\theta=\theta_{M}$, is shown by the use of a black dotted line.}
\label{fig:figure7}
\end{figure}

In an attempt to better understand the origin of the bi-stability, we now consider the case of finite-amplitude temperature perturbations.
Starting from the diffusive base state for $Ra=8\times10^5$ and $\theta_M=0.8$ as before, the temperature perturbation $\theta'$ in the liquid phase (\textit{i.e.} for $z<h_0$) is chosen to be
\begin{equation}\label{equ:perturbation}
    \theta' =  A \; e^{-10\left(x-\lambda/2\right)^{2}} \sin\left(\frac{\pi z}{h_{0}}\right),
\end{equation}
where $A$ is the arbitrary amplitude of the perturbation, $\lambda$ is the length of the domain and $h_{0}=1-\theta_M=0.2$ is the initial fluid depth. 
This perturbation represents a localised temperature increase in the middle of the liquid phase.
The length of the domain, $\lambda$ is set to $3$.
The amplitude $A$ of the perturbation is then varied from infinitesimal values to finite values.
Figure \ref{fig:Figure_subcritical_bifurcation} shows the difference between the averaged height and its initial value for different values of $A$.
For small values of $A$ (typically $A\leq2$), the perturbation decays, as expected since the diffusive base state is linearly stable for this choice of parameters ($Ra_D<Ra_c$).
For larger values of $A$ however, the initial perturbation is able to locally melt the solid, therefore increasing the local fluid depth so that the local effective Rayleigh number becomes supercritical and convection can be sustained.
This initially local patch of convective fluid spreads slowly throughout the domain.
This is best seen in wide horizontal boxes to minimise confinement issues, as seen in figure \ref{fig:figure7} where we increased the aspect ratio to $\lambda=6$. 
The speed at which this local convection patch propagates in the thermally-stable fluid can be estimated directly by calculating the slope from the dashed lined in figure \ref{fig:figure7}(a).
It is of the same order as the vertical diffusion time used to re-scale our equations.
This indicates that the heat diffusion in the solid dictates the speed at which the propagation occurs.

\subsection{Discussion \label{sec:discuss}}

\begin{figure}[t!]
    \centering
    \includegraphics[height=0.32\textwidth]{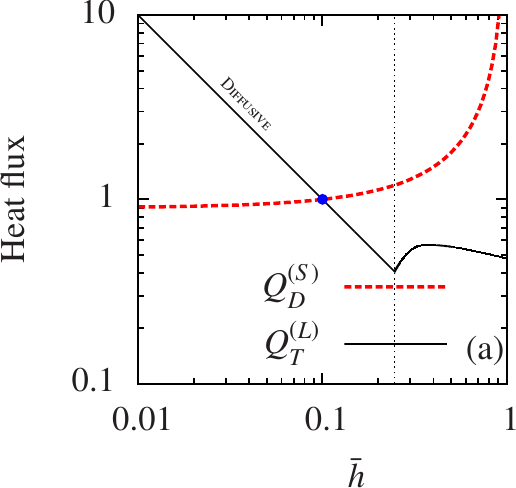}
    \hfill
    \includegraphics[height=0.32\textwidth]{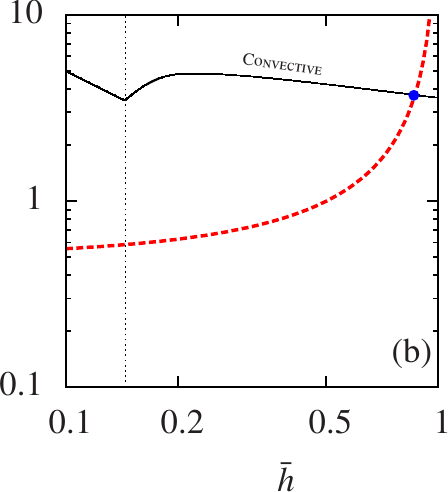}
    \hfill
    \includegraphics[height=0.32\textwidth]{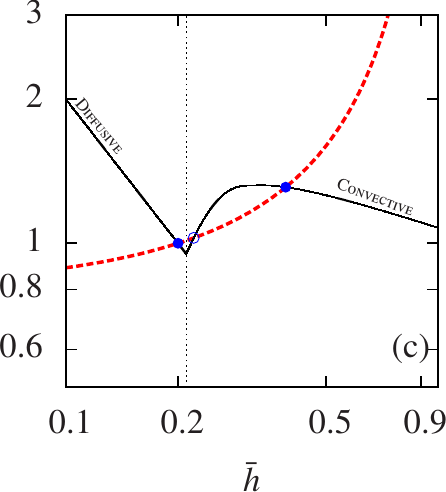}

    \caption{Heat fluxes across each layer of the system as a function of the averaged fluid layer depth $\overline{h}$. $Q_D^{(S)}$ is the diffusive flux through the solid layer defined by equation~\eqref{eq:heatflux_solid_diff}. $Q_T^{(L)}$ is the total heat flux through the liquid layer from our simple model for the Nusselt number including diffusive ($Nu=1$), weakly non-linear ($Nu\approx Ra_e$) and turbulent ($Nu\approx Ra_e^{1/4}$) regimes. Full and empty circles correspond to stable and unstable equilibria respectively. The dotted vertical line corresponds to the critical height above which convection sets in. From left to right, the cases correspond to the results discussed in Figures~\ref{fig:figure3}, \ref{fig:figure4} and \ref{fig:figure5} respectively.}
    \label{fig:heat_fluxes}
\end{figure}

The existence of the bi-stability has been discussed in the previous sections by either varying the initial position of the solid-liquid interface or by using a finite amplitude perturbation of a diffusive stable state.
We now discuss the origin of this bi-stable regime and whether it exists for all values of $\theta_{M}$ and $Ra$.
We recall that in section \ref{sec:with_convection}, we assumed that the Nusselt number was only function of the effective Rayleigh number far from the threshold of the thermal convection instability in the liquid.
However, we need a more refined model valid for any values of the effective Rayleigh number since bi-stability occurs near the threshold of the convective instability.
In an attempt to do so, we define the normalised distance from the onset of convection by $\zeta=(Ra_{e}-Ra_{c})/Ra_{c}$ and look for a general law $Nu(\zeta)$.
We consider the diffusive ($\zeta<0$), the weakly nonlinear ($0<\zeta<1.3$) and the fully nonlinear regimes ($\zeta>1.3$).
Hence, a continuous piece-wise model is obtained for the Nusselt number for any value of $\zeta$, and further details can be found in Appendix~\ref{sec:nusselt}.
Note that the following conclusions do not qualitatively depend on these particular choices.
The model is able to predict the existence of bi-stability provided the transition from weakly-nonlinear to fully nonlinear regimes is included.
The underlying assumption of our model is that the convectively-unstable flow below the interface behaves similarly to classical Rayleigh-B\'enard convection at all times, even when the system is out-of-equilibrium.
This has indeed been observed previously \citep{esfahani2018basal,favier_purseed_duchemin_2019} (see also appendix \ref{sec:nusselt}) and assumes a time-scale separation between the fluid motion and that of the interface (this is justified except in the low Stefan number limit).
Figure \ref{fig:heat_fluxes} shows the diffusive heat flux in the solid given by equation (\ref{eq:heatflux_solid_diff}) and the total heat flux in the liquid from this model.
Results are shown as a function of the average fluid depth for the three cases discussed in section \ref{sec:initial_cond}.
The averaged fluid depth $\bar h$ is systematically varied such that any intersection between the two curves corresponds to an equilibrium position $h_\infty$, solution of equation~\eqref{eq:fluxbalance}.

The prediction of the model for the case discussed in Figure~\ref{fig:figure3} is represented in the left panel of Figure~\ref{fig:heat_fluxes}. 
We recall that for this case, we considered $\theta_{M}=0.9$ and $Ra=10^{6}$. 
Only one intersection exists for this particular case so, for any initial value of $\bar h$, the system will converge to the corresponding stable equilibrium.
Since this intersection occurs in the diffusive branch of the total heat flux across the liquid, the nature of this equilibrium is diffusive and $h_\infty=1-\theta_{M}$, as expected.
This is consistent with the results of Figure~\ref{fig:figure3}(a).
The middle panel of Figure~\ref{fig:heat_fluxes} depicts the case discussed in Figure~\ref{fig:figure4} where $\theta_{M}=0.5$ and $Ra=10^{6}$. 
In this particular case, for all values of $\bar h$, only one equilibrium exists at the intersection between the convective branch of the heat flux across the liquid and the diffusive heat flux across the solid.
This is again consistent with the results of Figure~\ref{fig:figure4}(a).
Finally, the bistable case is illustrated in the right panel of Figure~\ref{fig:heat_fluxes} which corresponds to $\theta_{M}=0.8$ and $Ra=8\times 10^{5}$ (see Figure~\ref{fig:figure5}).
In that case, there are three intersections and thus three possible equilibria. 
The first equilibrium is a typical diffusive equilibrium at $h_\infty=1-\theta_{M}$ while the last is a far from threshold convective equilibrium.
Those are the two stable solutions observed in Figure~\ref{fig:figure5}.
The intermediate unstable equilibrium has not been observed in our simulations and thus separates the two basins of attraction of the other two stable solutions.

\begin{figure}[!t]
\centering
        \begin{subfigure}[l]{0.48\textwidth}
         \centering
         \includegraphics[width=1.05\textwidth]{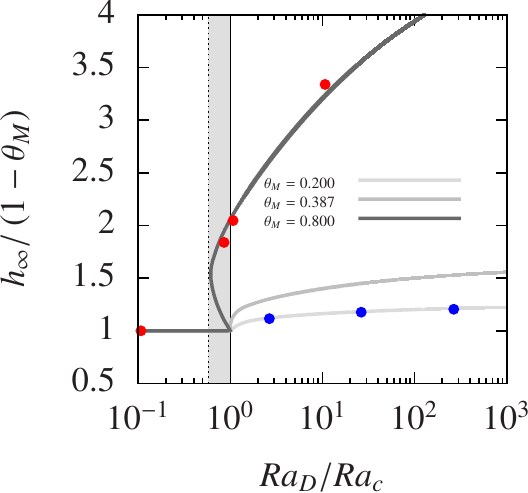}
         \caption{}
        \label{fig:hversusRat}
     \end{subfigure}
     \hfill
     \begin{subfigure}[r]{0.48\textwidth}
         \centering
         \includegraphics[width=0.98\linewidth]{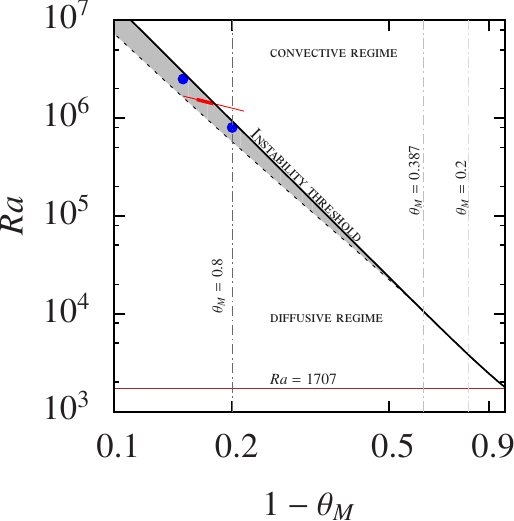}
         \caption{}
        \label{fig:phase_diagram}
     \end{subfigure}
    \caption{ (a) Normalised equilibrium height as a function of a normalised Rayleigh number for three different melting temperatures corresponding to the three vertical lines in Figure~\ref{fig:phase_diagram}. We recall $Ra_{D}= Ra \times \left(1-\theta_{M}\right)^{4}$. (b) A $\left(\theta_{M},Ra\right)$ phase diagram. The blue points represent the numerical points for which bi-stability has been observed. The red line represents the experimental domain investigated by M\"uller \& Dietsche (1985) where the thick segment denotes the range for which they found bi-stable solutions.}
\label{fig:figure9}
\end{figure}

We now ask whether this bistability regime exists for all pairs of control parameters $\left(\theta_{M}, Ra\right)$?
We address this question by solving for the equilibrium height $h_{\infty}$ using the flux balance given by equation~\eqref{eq:fluxbalance}.
The Nusselt number is estimated using the model discussed in Appendix~\ref{sec:nusselt}. 
Figure~\ref{fig:hversusRat} shows the normalised equilibrium heights as a function of a normalised Rayleigh number for three distinct melting temperatures.
The continuous lines correspond to the numerical solutions of equation~\eqref{eq:fluxbalance} for $\theta_{M}=0.2$, $0.387$ and $0.8$ evaluated over a wide range of $Ra$.
The blue and red dots correspond to numerical data obtained for $\theta_{M}=0.2$ and $\theta_{M}=0.8$ respectively.
In addition to the good agreement between the model prediction and the simulations (shown as full symbols), we also see a multiple solution domain appearing for $\theta_{M}=0.8$ (grey region in Figure~\ref{fig:hversusRat}).
In this particular case, for a small range of $Ra$, three equilibria are possible.
We then solve this equation for a wide range of control parameters, and the greyed area in Figure~\ref{fig:phase_diagram} represents the values of $\theta_{M}$ and $Ra$ for which three solutions are possible.
The continuous black line in Figure~\ref{fig:phase_diagram} represents the convection instability threshold above which only convective equilibria are possible.
The grey area again corresponds to the control parameters for which the model predicts multiple solutions.
This bi-stable regime exists in a wide band below the threshold but eventually disappears for values of $\theta_M<0.387$.
The limit $\theta_M\rightarrow0$ indeed corresponds to standard Rayleigh-B\'enard convection since the liquid layer has a negligible thickness and does not affect the dynamics (we recover that when $\theta_M\rightarrow0$, $h_\infty\rightarrow1$, $Ra_e\approx Ra$ and the threshold occurs at $Ra=1707$ as expected, see the horizontal line in Figure~\ref{fig:phase_diagram}).
We also report in Figure~\ref{fig:phase_diagram} our numerical simulations where bi-stability was observed by systematically varying the initial liquid depth and checking that two stable solutions are reached after several diffusive timescales (as in Figure~\ref{fig:figure5}(a)).
Note that exploring the bi-stable regime systematically via numerical simulations is a demanding task, since many simulations have to be run for several diffusive timescales for each set of control parameters.
Finally, the experimental results of \cite{dietsche1985influence} are reproduced using our dimensionless units.
Using well-controlled experiments, they observed bi-stability over a wide range of parameters indicated by the thick red line.
Note that our simple model overestimates the range of parameters for which bi-stability is observed compared to the experimental results.
This can be attributed to several differences between our idealised model and the experiment (\cite{dietsche1985influence} used a high Prandtl number fluid $Pr=17$ compared to $Pr=1$ used in the present study) but we suspect the main source of uncertainties is related to the presence of the non-planar topography typical of the convective solutions.
In particular, the diffusive heat fluxes have been derived neglecting the topography and a more refined analysis (following for example the perturbative approach of \cite{favier_purseed_duchemin_2019}) is probably required to more accurately predict the disappearance of the bi-stable convective branch.
This is particularly true in the limit $\theta_M\rightarrow1$ for which the liquid depth is small and the topography cannot be neglected.
This remains to be further analysed in future studies.

\section{Conclusion\label{sec:level10}}

We performed 2D direct numerical simulations of a liquid layer bounded by two fixed-temperature horizontal plates.
When the melting temperature of the pure substance is comprised between the upper and lower temperatures, a phase-change interface lies inside the domain.
We have shown that, depending on the control parameters, this system exhibits equilibrium states that can be of a conductive or a convective nature.
The conductive equilibrium can be described as a planar interface separating the idle liquid phase from the solid one.
On the other hand, the convective equilibrium happens when the Rayleigh number based on the fluid layer height is large enough, causing the liquid to convect.
The melting front stops when the heat fluxes in both phases are balanced, leading to a convective equilibrium with convection rolls and a non-planar interface. 

Assuming that the convection below the interface behaves as standard Rayleigh-B\'enard convection allowed us to predict the mean equilibrium position of the interface.
This approach is in good agreement with our numerical simulations over a wide range of melting temperatures and Rayleigh numbers.
In marginal cases, when the static equilibrium is close to being unstable to Rayleigh-B\'enard convection, we observed bi-stable states, for which both convection and diffusion equilibria are observed for the same control parameters.
This new sub-critical convective branch can be obtained by perturbing the diffusive equilibrium with a finite amplitude perturbation, leading to local convection, which eventually invades the whole domain.
The final convective equilibrium is reached when the diffusive heat flux across the receding solid layer balances the convective heat flux in the fluid.

We recall that the Prandtl number was fixed to unity throughout our study.
It is however well-known that the heat flux carried by Rayleigh-B\'enard convection depends on this dimensionless number \citep{grossmann_lohse_2000}.
Hence, one would need to factor in this parameter in the heat flux scaling, \textit{i.e.} using a more general scaling of the form $Nu=f(Ra,Pr)$ instead of equation~\eqref{eq:Nusselt_scaling}, to get an accurate equilibrium prediction.
Similarly, the Stefan number is also fixed at unity throughout the paper.
Since the Stefan number only affects the transient melting or solidifying phases (it only appears in the Stefan condition~\eqref{eq:st2} in factor of the interface velocity), we do not expect that this parameter will affect the equilibrium height.
Our theoretical model leading to equation~\eqref{eq:fluxbalance} does not depend on the Stefan number for example.
We have checked numerically that, for intermediate values of the Stefan number $10^{-1}\leq St\leq 10$, the equilibrium height is unchanged.
The transient solidification or melting phases are of course affected, and become longer as the Stefan number increases, but the asymptotic equilibrium height is independent of the Stefan number.
The impact of the Stefan number on the bi-stable regime is less obvious however.
For very low Stefan numbers, phase change processes could be so fast as to prevent the growth of thermal convection irrespective of initial conditions, thus only leading to a diffusive equilibrium.
It is however likely that the bi-stable regime observed here for $St=1$ subsists in the large Stefan limit, although this remains to be explored in details.
Note also that our model is not applicable to water since the latter has a maximum density at $4$ degrees Celsius.
This can lead to a thermally-stratified layer near the interface where the convecting part of the liquid would interact with the stratified layer rather than directly with the solid-liquid interface \citep{couston2017dynamics}.
This in turn can possibly affect the melting and solidification processes along with the equilibrium states discussed in the present study.

A generalisation of our 2D results to three dimensions would be interesting and a better comparison to the experimental works of \cite{davis_muller_dietsche_1984}.
Such simulations have recently been realised in the case of an isothermal solid \citep{esfahani2018basal}.
Extending their results to the case of a solid layer cooled at a temperature below the melting temperature would be valuable.
While it is known that Rayleigh-B\'enard convection can significantly differ between 2D and 3D dimensions \citep{vanderpoel_stevens_lohse_2013}, we nevertheless expect our approach to remain valid provided that one takes into account the possible change in heat flux through the Nusselt-Rayleigh scaling.

Finally, the bi-stable regime observed in this paper deserve a more detailed analysis. 
The propagation of the convective motions into the stable diffusive region could be characterised as a percolation mechanism \cite{POMEAU19863}.
In addition, the simultaneous existence of both quiescent fluid and convective motions is similar to other convective systems where bi-stability and spatially-localised states are observed \citep{Knobloch_2008}.
This is the case for example of magnetoconvection \citep{BLANCHFLOWER199974}, binary-fluid convection \citep{batiste_knobloch_alonso_mercader_2006}, double-diffusive convection \citep{bergeon2008} or rotating convection \citep{beaume_bergeon_kao_knobloch_2013}.
Whether such stable localised states can exist in the current system involving liquid-solid phase change remains to be confirmed. 

\acknowledgments{Centre de Calcul Intensif d'Aix-Marseille is acknowledged for granting access to its high performance computing resources.}

\appendix
\begin{section}{Phase-field equations and numerical parameters}\label{sec:level11}

The numerical method used in this paper is similar to the method used in \citep{favier_purseed_duchemin_2019} where they solve the physical problem described in section~\ref{sec:level2} using a phase-field approach.
A continuous order parameter $\phi\left(x,z,t \right)$ takes the values zero and unity in the solid and liquid phases respectively.
This results in a continuous interface where $\phi \in ]0,1[$ over a width $\epsilon$.
The phase-field equation associated to this particular problem is given by\citep{Wang1993} 
\begin{equation}
\label{eq:pf}
\frac{\epsilon^2}{m}\frac{\partial\phi}{\partial t}  =  \epsilon^2\nabla^2\phi + \frac{\alpha\epsilon}{St} \left(\theta-\theta_M\right)\frac{dp}{d\phi} - \frac{1}{4}\frac{d g}{d\phi},
\end{equation}
where $g(\phi)=\phi^2(1-\phi)^2$ and $p(\phi)=\phi^3(10-15\phi+6\phi^2)$ are functions which ensure that the phase-field is either zero or unity everywhere except close to the solid/liquid interface.
The particular choice of these functions results from thermodynamical considerations \citep{Wang1993,favier_purseed_duchemin_2019}.
$\epsilon$, $m$ and $\alpha$ denote the interface width, the mobility and the coupling parameter between the phase-field and the temperature field respectively.
The Stefan problem described in the main text is asymptotically recovered in the double limit $\epsilon\ll1$ and $\alpha> St/\epsilon$ \citep{Caginalp1989,Wang1993} while the mobility is fixed to unity.
Following the convergence study presented in \cite{favier_purseed_duchemin_2019}, the width of the interface is chosen to be close to the maximum grid spacing, while the coupling parameter is given by $\alpha\gtrsim\epsilon^{-1}$.
We additionally solve the heat equation and the Navier-Stokes equations under the Boussinesq approximation
\begin{eqnarray}
\label{eq:tempeq_adim}
\frac{\partial\theta}{\partial t} +\bm{u}\cdot\nabla \theta & = &
\nabla^2\theta-St\frac{dp}{d\phi}\frac{\partial\phi}{\partial t} \ , \\
\label{eq:mompen}
\frac{1}{Pr}\left(\frac{\partial\bm{u}}{\partial t}+\bm{u}\cdot\nabla\bm{u}\right) & = & -\nabla P+ {\mathrm Ra}\, \theta \, \bm{e}_z+\nabla^2\bm{u}-\frac{ f(\phi) \bm{u}}{\eta} \ .
\end{eqnarray}
The last term in equation~\eqref{eq:tempeq_adim} corresponds to latent heat effects.
An immersed boundary method called the volume penalisation \cite{Angot1999} is used to ensure no-slip boundary condition at the interface. The last term in equation~\eqref{eq:mompen} is the penalisation term and ensures an exponential decay of the velocity in the solid provided $\eta$ is small enough.
The results discussed in the main paper were obtained with a mask function $f(\phi)=1-\phi$.
Although this choice is rather arbitrary (any function continuously varying from $0$ in the liquid phase to $1$ in the solid is appropriate), we have checked that the results discussed in this paper do not depend on this arbitrary choice.
The function $f(\phi)=(1-\phi)^{2}$ was for example used in \citep{favier_purseed_duchemin_2019} and we have checked that the nature of the solution we obtained (convective or diffusive) is the same for this other mask function.
The relative error on the equilibrium height $h_{\infty}$ depending on the mask function used does not exceed $1\%$. 
$\eta$ is the penalisation parameter and must be small enough to model no-slip boundary conditions on the solid/liquid interface.
Here, and following the recent work of \cite{hester2019improving}, we choose the approximate scaling $\eta\lesssim\epsilon^2$, while ensuring that $\eta$ is larger than the time step for stability reason.
We note that an extended asymptotic analysis must be performed to ensure that second order convergence (as discussed in \cite{hester2019improving}) with respect to the penalisation parameter is indeed achieved in our configuration involving buoyancy forces.

Most of the simulations described in the main text have been performed using the same numerical approach as in \cite{favier_purseed_duchemin_2019}.
For comparison, some of the cases have been solved by using the open-source pseudo-spectral code Dedalus \citep{DEDALUS}.
We use Chebyshev polynomial functions in the $\bm{e}_{z}$ direction and a Fourier decomposition in the periodic $\bm{e}_{x}$ direction.
A fourth-order Runge-Kutta scheme is chosen for time integration.
For the exact same model and physical parameters, an excellent agreement between the two numerical solvers is obtained, with a relative error on the equilibrium height around $10^{-6}$ for typical cases representative of the different regimes discussed in the paper. 

The numerical parameters for all cases studied in this paper are given in Table~\ref{table:param}.
Case A corresponds to the results discussed in figure \ref{fig:Main_figure}, case B to figures \ref{fig:figure3} and \ref{fig:figure4}, case C to figure \ref{fig:figure5}, case D to \ref{fig:Figure_subcritical_bifurcation} and finally case E to figures \ref{fig:ht_plot_subcritical} and \ref{fig:Visualisation}.

\begin{table}[ht]
\begin{center}
\begin{tabular}{ |m{1.5cm}|m{1.5cm}|m{1.5cm}|m{1.5cm}|m{1.5cm}|m{1.5cm}|m{1.5cm}|m{1.5cm}| } 
 \hline
 Case & $N_{x}$ & $N_{z}$ & $\epsilon$ & $\alpha$ & $\lambda$ & $\eta$ \\ 
 \hline
 A & 512 & 256 & $2\times10^{-3}$ & 500 & 6 & $1.2\times10^{-6}$ \\ 
 B & 512 & 512 & $4\times10^{-3}$ & 1250 & 9 & $3\times10^{-7}$\\
 C & 1024 & 512 & $3\times10^{-3}$ & 667 & 8 & $9\times10^{-7}$\\ 
 D & 512 & 512 & $2\times10^{-3}$ & 1500 & 3 & $4\times10^{-7}$\\
 E & 1024 & 512 & $2\times10^{-3}$ & 1500 & 6 & $4\times10^{-7}$\\
 \hline
\end{tabular}
\end{center}
\caption{List of numerical parameters for all simulations described in this paper.}
\label{table:param}
\end{table}
\end{section}

\begin{section}{Model for the Nusselt number}\label{sec:nusselt}

In this appendix, the simplified model for the convective heat flux is detailed.
We assume that the convection is responding instantaneously to any change in topography and that it behaves as standard Rayleigh-B\'enard.
Physically, this is justified when the variation of topography is slow compared to the fluid turnover time, \textit{i.e.} when the Stefan number is large.
We define $\zeta$ as a normalised distance from the threshold: 
\begin{equation}
    \zeta = \frac{Ra_{e}-Ra_{c}}{Ra_{c}} \ ,
\end{equation}
where we recall that the critical Rayleigh number is a function of $\theta_M$ \citep{davis_muller_dietsche_1984}.
Based on this parameter, we can define three distinct regimes: diffusive, near threshold convection and far from threshold convection. The relationship between this parameter and the Nusselt number is chosen as follows: 
\[
    Nu = \begin{cases}
            = 1 \quad \text{if} \quad \zeta \leq 0 \\
            = 1 + \upsilon \zeta \quad \text{if} \quad  0 < \zeta \leq 1.3\\
            = \delta \zeta^{\beta} \quad \text{if} \quad \zeta > 1.3
            \end{cases}
\]
where we choose the following arbitrary values $\upsilon=0.88$, $\delta=0.27\times Ra_{c}^{\beta}$ and $\beta=0.25$.
These values are consistent with classical measurements of the Nusselt number close to threshold \citep{Chandra1961}.
The transition between the near threshold convection and the far from threshold convection is then smoothed by using a third-order polynomial interpolation from $\zeta=0.1$ to $\zeta=2.5$. 
Figure~\ref{fig:Nusselt_Rayleigh} shows the Nusselt number as a function of the effective Rayleigh number.
The model corresponds to the dotted black line (which is only shown for $\theta_{M}=0.9$ for clarity, the other values of $\theta_M$ being nearly indistinguishable on this log-log representation). 
From the simulations of the equilibrium states discussed in section~\ref{sec:level3}, their respective Nusselt number at equilibrium and effective Rayleigh number at equilibrium are plotted by the use of red dots.
We see a good agreement between the numerical data and our model.

\begin{figure}[ht!]
    \centering
    \includegraphics[width=0.6\linewidth]{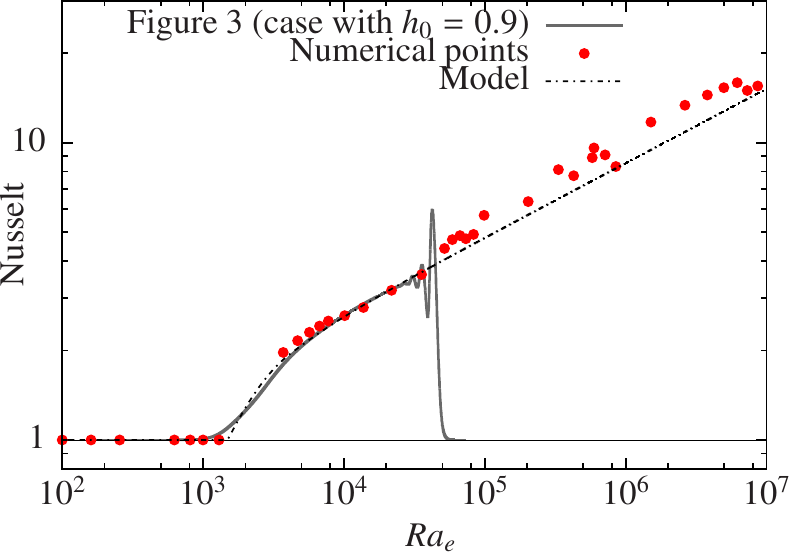}
    \caption{Nusselt number as a function of the effective Rayleigh number. The black line corresponds to the model, the red dots represent the Nusselt number and effective Rayleigh number at equilibrium for all the simulations in section \ref{sec:level3}. Finally, the grey line shows the transient evolution of the Nusselt number as a function of the effective Rayleigh number for the case discussed in Figure~\ref{fig:visus_fig3}  ($h_{0}=0.9$).}
    \label{fig:Nusselt_Rayleigh}
\end{figure}

We can now test the validity of our quasi-static assumption by considering a transient case where the average fluid depth evolves with time. 
For example, the case discussed in Figure~\ref{fig:figure3} where $h_{0}=0.9$ and $\theta_{M}=0.9$. 
The grey line in Figure~\ref{fig:Nusselt_Rayleigh}, represents the evolution of the instantaneous Nusselt number as a function of the instantaneous effective Rayleigh number during the solidification process. 
Note that this curve should be read from right to left.
Initially, the Nusselt number is unity since we initialise our simulation with a linear temperature profile with small perturbations but quickly increases since $Ra_{e} \gg Ra_{c}$.
As the solid phase grows, the effective Rayleigh number decreases which in turn decreases the Nusselt number, until the diffusive state is reached. 
This case study is in good agreement with our model for all values of the effective Rayleigh number.
A small mismatch is observed near the threshold which can be attributed to the presence of the topography.
The classical super-critical bifurcation indeed becomes imperfect when the boundary is not exactly horizontal \citep{Kelly1978}.
This is a first indication that the presence of a topography plays an important role on the Nusselt number (particularly near the threshold) and consequently the heat fluxes. 
This can be one of the reasons why our model overestimates the range of parameters for which bi-stability is observed.
Note finally that our model underestimates the Nusselt number at large Rayleigh numbers, which is again a consequence of the back-reaction of the topography on the flow, as discussed in~\cite{favier_purseed_duchemin_2019}.
\end{section}

\bibliographystyle{apalike}
\bibliography{biblio}

\end{document}